\documentclass[10pt]{article}

\pdfoutput=1
\usepackage[pdftex]{color}
\usepackage{amssymb}
\usepackage{amsthm} 
\usepackage{amsmath}
\usepackage{latexsym}
\usepackage{amscd}
\usepackage{graphicx}
\usepackage[pdftex, colorlinks=true, citecolor=green]{hyperref}
\usepackage{lscape}
\usepackage{multirow}
\usepackage{wrapfig}

\newcommand{\ds}{\displaystyle}

\newcommand{\ben}{\begin{equation}}     
\newcommand{\eeqn}{\end{equation}}
\newcommand{\bey}{\begin{eqnarray}}
\newcommand{\eey}{\end{eqnarray}}

\newtheorem{thm}{Theorem}[section]

\newtheorem{defn}[thm]{Definition}

\begin{document}

\begin{flushleft}
{\Large
\textbf{Computing brain networks with complex dynamics}
}
\\
\vspace{4mm}
Anca R\v{a}dulescu$^{*,}\footnote{Assistant Professor, Department of Mathematics, State University of New York at New Paltz; New York, USA; Phone: (845) 257-3532; Email: radulesa@newpaltz.edu}$, Johan Nakuci$^2$, Simone Evans$^3$, Sarah Muldoon$^4$
\\
\indent $^1$ Department of Mathematics, SUNY New Paltz
\\
\indent $^2$ School of Psychology, Georgia Institute of Technology
\\
\indent $^3$ Department of Computational Neuroscience, Dartmouth College
\\
\indent $^4$ Department of Mathematics, University at Buffalo
\end{flushleft}

\begin{abstract}
One important question in neuroscience is how global behavior in a brain network emerges from the interplay between network connectivity and the neural dynamics of individual nodes. To better understand this theoretical relationship, we have been exploring a simplified modeling approach in which we equip each node with discrete quadratic dynamics in the complex plane, and we study the emerging behavior of the resulting complex quadratic network (CQN). The long-term behavior of CQNs can be represented by asymptotic fractal sets  with specific topological signatures going far beyond  those described in traditional single map iterations. 

In this study, we illustrate how topological measures of these asymptotic sets can be used efficiently as comprehensive descriptors and classifiers of dynamics in tractography-derived connectomes for human subjects. We investigate to what extent the complex geometry of these sets is tied to network architecture (on one hand) and to the network behavior (on the other). This helps us understand the mechanics of the relationship between the subject's brain function, physiology and behavior and their underlying connectivity architecture.
\end{abstract}

\section{Introduction}

Many natural systems can be described as collective behavior of units that are part of a larger network~\cite{benner2014large}. To gain insight into how the ensemble behavior of these units operates in natural systems, dynamical systems modeling has been widely used to understanding the associated nonlinear and chaotic phenomena~\cite{porter2016dynamical}. Modeling each coupled unit in a fashion that captures 
natural dynamics with accuracy is important for applications. Unfortunately, accuracy in modeling comes with a price, and matching the complexity of real systems in a mathematical model leads to increased analytical and computational challenges. This often compromises tractability of the model behavior, and takes away from its potential applicability. 

To bridge this gap, we have developed a simplified mathematical model that is accessible analytically and computationally, and that captures both connectivity and coupled dynamics in terms of discrete iterations in the complex plane, in a canonical framework with increased prediction power stemming from this combination. Here we describe the model and develop a data-driven application to classifying ensemble dynamics in a brain tractography data set.

\subsection{Graph theory in networks}

Network science has been focused on the graph theoretical aspect of networks, particularly on investigating the relationship between the network's connectivity architecture and its function. By applying graph theoretical measures, one can investigate for example the sensitivity of a system's temporal behavior to removing/adding nodes or edges at different places in the network structure. This is true in particular for the brain, which can be viewed as a ``dynamic network,'' self-interacting in a time-dependent fashion at multiple spatial and temporal scales, to deliver an optimal range for biological functioning. Graph theoretical approaches have been applied to brain networks to understand organizational and functional neural principles~\cite{bullmore2009complex,sporns2011human,sporns2011non}, with results supporting certain architectural and topological properties, such as modularity, small-worldness, the existence of hubs and ``rich clubs''~\cite{sporns2022structure,park2013structural}. Such measures have been used successfully on brain imaging data, tying pathological behavioral patterns to specific abnormalities in connectivity~\cite{fornito2017opportunities,korgaonkar2014abnormal,erdeniz2017decreased}. 

However, graph theoretical measures cannot explain in and of themselves the mechanisms by which connectivity patterns act to change systemic behavior. Therefore, a network model needs to incorporates both graph structure and node-wise dynamics in one unified framework, which can then be used to interpret empirical results and make predictions. To this aim, a lot of effort has been invested towards formal modeling approaches that would explain how brain connectivity patterns may affect its functional dynamics (from biophysical models~\cite{gray2009stability} to simplified systems~\cite{siri2007effects}). Navigating the trade-off between faithful representation and mathematical simplicity can be difficult. In computational neuroscience in particular, models of brain dynamics often lie at either end of the spectrum: incorporating intractable complexities in favor of biological realism, or sacrificing accuracy (and with it, practical impact) in favor of mathematical simplicity. Indeed, it has become increasingly clear that the construction of a realistic, data-compatible computational model presents many difficulties related to dimensionality, computational cost, addressing multiple scales, and even simple bookkeeping. These difficulties make it nearly impossible to deliver any useful general results relating brain connectivity patterns to brain dynamics and observed behavior.

We developed a framework that addresses the limitation in the current network models investigating the structure-to-function relationship in a canonical and tractable mathematical framework. Here, we show how we can use this framework to place and solve outstanding questions in dynamic networks. 

The model relies on ideas from traditional complex quadratic dynamics to address questions which are notoriously difficult in realistic models of natural networks. By rephrasing these questions in terms of complex map iterations presents an approachable way that can lead to novel results, as well as to progress on applications. Complex plane iterations represent good candidates as building blocks for this framework because their dynamics have been thoroughly studied, are well understood, and can be illustrated geometrically by sets with extraordinarily complex, but topologically quantifiable structure. These structures can be used to describe and understand the long-term dynamics of the system, and can be harnessed to generate the highly needed classification and prediction tools based on network dynamics, rather than solely on graph theoretical measures.

\subsection{Complex quadratic dynamics in single maps}

Dynamic behavior of iterated complex quadratic maps refers to the behavior of sequences in the complex plane obtained by picking an initial point $z_0 \in \mathbb{C}$, and repeatedly applying a function in the family $f_c(z) = z^2+c$ (for a fixed value of the parameter $c \in \mathbb{C}$). This way, each complex number $z_0$ has an associated ``orbit'' under the function $f_c$:
$$z_0 \to f_c(z_0) \to f_c(f_c(z_0)) \to \hdots \to f_c^{\circ n}(z_0) \to \hdots$$

\noindent For each value of the parameter $c$, the corresponding function $f_c$ presents with different orbits, with a wide range of behaviors (which may vary from trivial to very complicated, depending on both the parameter $c$ and the initial condition $z_0$ itself). The study of these systems emerged historically at the start of the 20th century, with the work of Fatou~\cite{Fatou} and Julia~\cite{Julia}. These simple iterative rules describe rich, complex systems, and resulting objects with intricate fractal geometry. Despite the seemingly elementary and natural essence of the questions on orbit behavior for quadratic maps, investigating them required the development of a new field of mathematics, encompassing real and complex analysis, topology and computational methods. 


\begin{figure}[h!]
\begin{center}
\includegraphics[width=\textwidth]{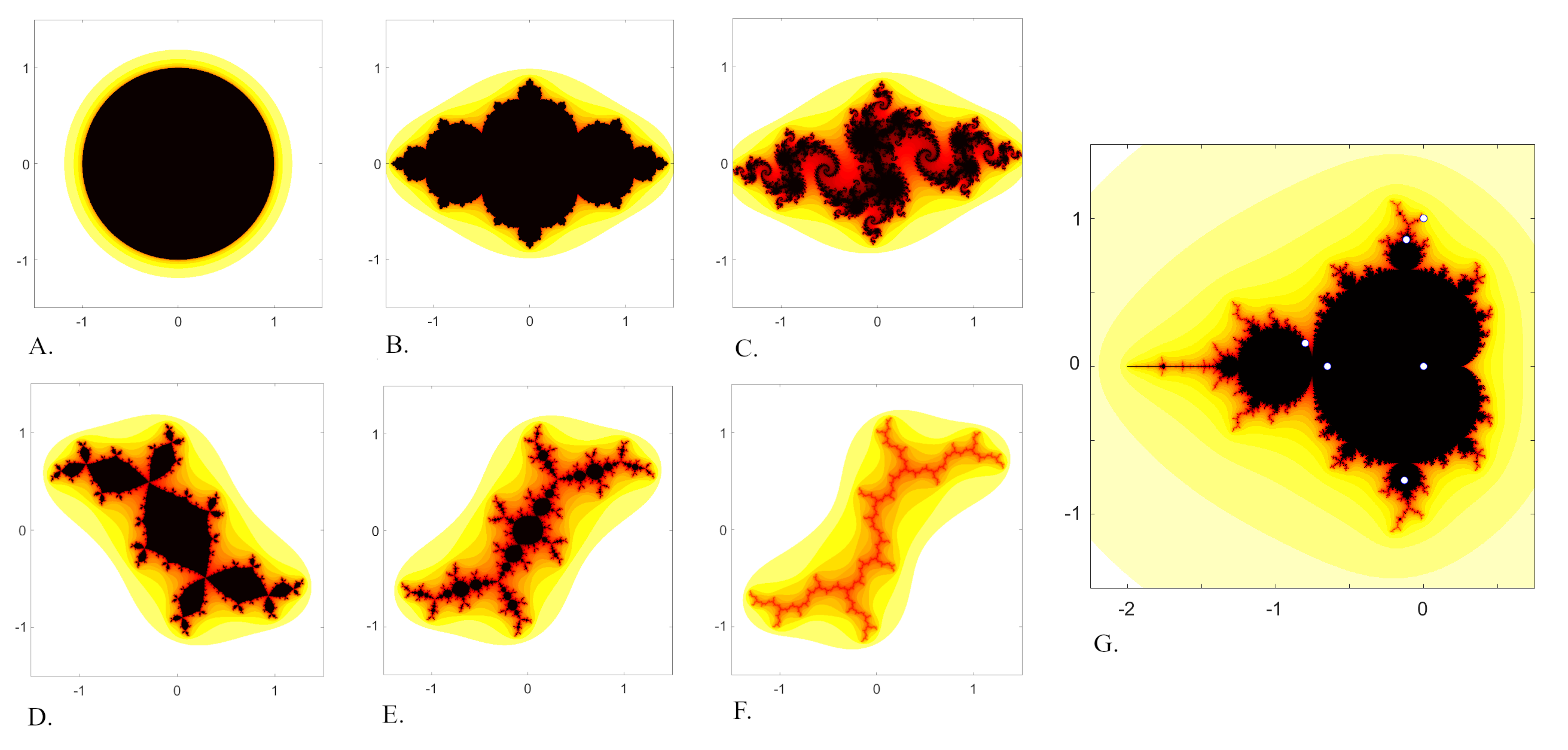}
\end{center}
\caption{\emph{{\bf Example asymptotic sets for single map iterations.} 
The left panels illustrate the Julia set corresponding to the following parameter values: {\bf A.} $c=0$ (unit circle); {\bf B.} $c=-0.65$; {\bf C.} $c=-0.8+0.156i$; {\bf D.} $c=-0.13 + 0.77i$; {\bf E.} $c=-0.117+0.856i$; {\bf F.} $c=i$ (dendritic set). The prisoner set is shown in black; the colors represent how fast (in how many iterations) escapees fall outside the disc of radius 2. {\bf G.} Traditional Mandelbrot set, illustrating (in black) the values of $c$ for which the Julia set is connected, and for which critical point $z_0=0$ is a prisoner. The colors outside of the set represent how fast $z_0=0$ iterates out of the disc of radius 2.}}
\end{figure}

Complex quadratic iterations of individual functions in the family $f_c = z^2+c$ deliver rich dynamics, and provide textbook recipes for creating fractal asymptotic sets, with history going back more than a century~\cite{Fatou,Julia}. For a parameter value $c \in \mathbb{C}$, the Julia set of the corresponding map $f_c$ is defined as the boundary between initial points $z_0$ whose orbits remain bounded under iterations of $f_c$ (prisoners), and those which escape to infinity under repeated iterations of the map (escapees). One interesting result refers to the existence of an \emph{escape radius} for these iterations: once the orbit transcends the radius $R=2$, the orbit is guaranteed to escape. The topological and fractal properties of Julia sets have been well studied, with major results relating the geometry of the Julia set with properties of the critical orbit of $z_0=0$~\cite{branner1992iteration,qiu2009proof,devaney2006criterion}. For iterations of single quadratic maps, it is known that the Julia set is either connected, if the orbit of the critical point $z=0$ is bounded, or totally disconnected, if the orbit of the critical point is unbounded. The postcritically bounded parameter locus is therefore the same as the Julia set connectedness locus in the parameter complex plane, and is known as the Mandelbrot set~\cite{branner1989mandelbrot}, the topology and properties of which have been extensively studied.  Although establishing relationships between quadratic dynamics and the topology of the Mandelbrot set has been highly nontrivial, many interesting properties have been established (such as connectedness and full Hausdorff dimension). What is remarkable in this context is that the behavior of the critical orbit of $z_0=0$ governs the potential behaviors of all the other orbits; hence the Mandelbrot set encompasses information on global system dynamics, which goes beyond simply describing the evolution of the system starting at ``rest''. For all these reasons, the Mandelbrot set remains both a staple in discrete dynamical systems, and a canonical representation for more general phenomena.

While other emerging areas of dynamical systems have been widely used towards understanding nonlinear and chaotic phenomena in natural systems, complex dynamics of quadratic maps have been thus far quite removed from such applications -- being considered a very theoretical field of study, difficult to tie to any realistic system. The possibility of coupling complex quadratic functions to study general dynamic principles in natural networks is, to the best of our knowledge, the first attempt at crafting applications for this mathematical framework. We explore extending this approach to the context of coupled networks, as a method to investigate how dynamics propagate in a network of given architecture, and to better understand how structural and functional connectivity are related.

\subsection{Network iterations, and applications to dynomics} 

While studying iterations of one single map at a time delivers tractability to the study of rich dynamics, realistic systems (as previously mentioned) are generally formed of many coupled units, and cannot be encoded by the dynamics of a single iterated map. In our work, we use quadratic iterations to understand the principles of how dynamic behavior emerges in large networks of nodes, and how it depends on the network structure. Each node is viewed symbolically as an integrator of internal and external inputs. We use the asymptotic behavior of multi-dimensional orbits (via the topological and fractal structure of Julia and Mandelbrot sets) to quantify dynamic behavior under perturbations of the network architecture. Topological landmarks of Mandelbrot sets provide valuable means of classification and comparison between different systems' behavior, as will be further summarized in the following section.

In our complex quadratic network (CQN) model, the node-wise dynamics is considered to be complex quadratic dynamics, within the family $f_c \colon \mathbb{C} \to \mathbb{C}$, $f_c(z) = z^2+c$. More specifically, in this framework, each network node adds all weighted inputs from adjacent nodes, and integrates the sum of inputs, in discrete time, as a complex quadratic map. Then the system takes the form of an iteration in $\mathbb{C}^n$:

\begin{eqnarray}
z_j(t) \longrightarrow z_j (t+1) &=& f_j\left(\sum_{k=1}^{n}{g_{jk} A_{jk} z_k} \right) \nonumber 
\label{mothermap}
\end{eqnarray}

\noindent where $n$ is the size of the network, $\ds A=(A_{jk})_{j,k =1}^n$ is the binary adjacency matrix of the oriented underlying graph, that is $A_{jk}=1$ if there is an edge from the node $k$ to the node $j$, and $A_{jk}=0$ otherwise. The coefficients $g_{jk}$ are the signed weights along the adjacency edges (in particular, $g_{jk}=0$ if there is no edge connecting $k$ to $j$, that is if $A_{jk}=0$). In isolation, each node $z_j(t) \to z_j(t+1)$, $1 \leq j \leq n$, iterates as a quadratic function $f_j(z) = z^2 + c_j$. When coupled as a network with adjacency $A$, each node will act as a quadratic modulation on the sum of the inputs received along the incoming edges (as specified by the values of $A_{jk}$, for $1 \leq k \leq n$). 

This ``network environment'' preserves, often in a weaker form, some of the properties and results determined in the traditional case of a single iterated quadratic map (e.g., the existence of an escape radius is guaranteed in some types of networks, but may fail in others). We used this to our advantage, and proceeded to study theoretically the effects of network architecture on its long-term dynamics. To do this, we needed to extend the definitions of asymptotic sets from the traditional context of single map iterations to the context of networks. Below, we will only provide the definitions and interpretations that lie within the scope of the current study.

\begin{defn}
We define the \textbf{multi-Mandelbrot set (or the multi-M set)} of the network the parameter locus of ${\bf c} = (c_1,...,c_n) \in \mathbb{C}^n$ for which the multi-orbit of the critical point $(0,...,0)$ is bounded in $\mathbb{C}^n$.
\end{defn}



\noindent Broadly speaking, one can interpret iterated orbits as describing temporal trajectories of an evolving system. Along these lines, we view a parameter ${\bf c}$ in the Mandelbrot set (for which the critical orbit that escapes to $\infty$) as representing a system with unsustainable long term dynamics when initiated from rest, while the ${\bf c}$ range for which the critical orbit remains bounded can be viewed as the ``sustainable dynamics locus.''

Since topological properties of $n$-dimensional complex sets are both difficult to visualize and hard to investigate, we started by considering Mandelbrot sets for networks with identical nodes ${\bf c} = (c,...c)$. These are object in the complex parameter plane $\mathbb{C}$, which we called equi-M sets: 

\begin{defn}
We call the \textbf{equi-Mandelbrot set (or the equi-M set)} of the network, the locus of $c \in \mathbb{C}$ for which the critical multi-orbit is bounded for \textbf{equi-parameter} ${\bf c} =(c,c,...c) \in \mathbb{C}^n$.  
\end{defn}

We test hypotheses tying connectivity to dynamic patterns using simple, low-dimensional networks, which are both analytically tractable and allow easier visualization and interpretation of the results. We found universal, unifying features: for example, all equi-M sets appear to exhibit a main cardioid, and some reminiscent bulb-like structure, although the traditional hyperbolic bulbs no longer exist, since network combinatorics must track behavior of many nodes simultaneously. Hence properties such as connectedness, size, position of the cusp and tail present with a lot of variability, based on the architecture of the underlying network.

\begin{figure}[h!]
\begin{center}
\includegraphics[width=\textwidth]{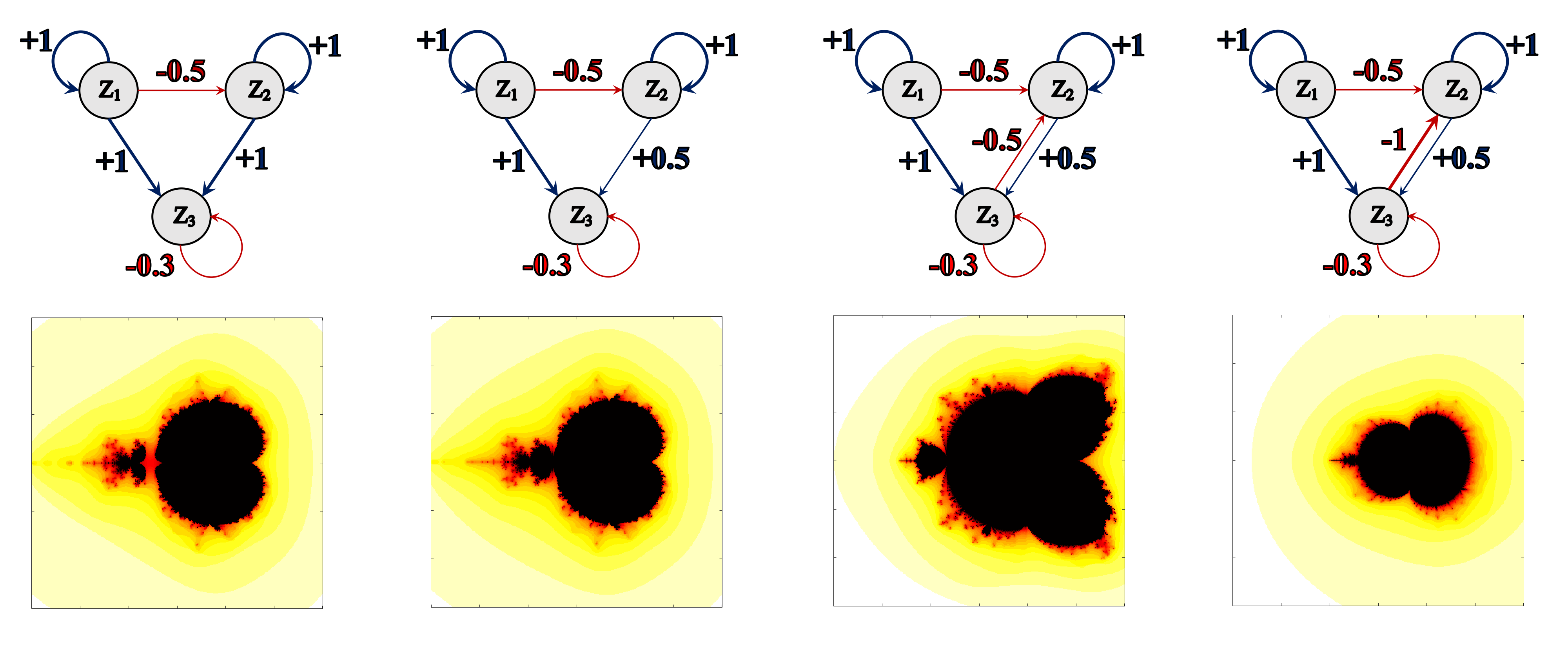}
\end{center}
\caption{\emph{\small{\bf Illustration of how perturbations in the network placement and weight affect the shape of equi-M sets.} Four networks are show in the top panels, with direction, sign and weight specified on each edge. The correponding equi-M sets are shown in the bottom panels. Changing to edge weights and adding negative feedback to the network induce quantifiable changes in the topology of the equi-M sets, altering the number of connected components, as well as the position and shape of the cusp.}}
\label{M_examples}
\end{figure}

We identified a few global network effects, associating larger connection weights with significantly smaller equi-M sets, and tying sparser networks to decreased node synchronization. We found that the presence of even weak inhibitory coupling is efficient in breaking down sets into connected components when introduced in a network formed of purely excitation connections. More interestingly, we also identified finer and less intuitive local effects, showing that changing the weight of one edge may lead to significant changes in the topology (Figure~\ref{M_examples}). This effect depends tightly on the position of the inserted/deleted edge in the network, and is strongly related to the contribution of the edge to graph-theoretical properties of the network, as well as to the weight of the edge.

 To fix these ideas with an example, Figure~\ref{M_examples} illustrates how a series of relatively small modifications to a 3-dimensional oriented network can have major consequences on the geometry of the corresponding equi-M set. In the network on the far left, nodes $z_1$ and $z_2$ feed with equal positive weights $g_{31}=g_{32}$ into the node $z_3$, producing a disconnected equi-M set (panel 1). When the right edge weight is weakened to $g_{32}=0.5$, the set becomes connected (panel 2). When introducing negative feedback $g_{32}=-0.5$ from the node $z_3$ to $z_2$, the cusp of the set explodes into two bulb-like structures (panel 3). These gradually disappear as the inhibitory feed back is further increased, so that at the point where $g_{32}=-1$, the pseudo-bulbs and the cusp are no longer visible (panel 4).

This shows that changes to the architecture and the weights of the connectome can induce the type of changes in network equi-M sets that are visually striking and topologically quantifiable~\cite{radulescu2017real,radulescu2019asymptotic}. This idea opens multiple areas of analysis. We are harnessing this possibility in a few potentially crucial directions. Here we focus on whether equi-M topology can be used to differentiate between different types of natural dynamic networks (e.g., between social and brain networks), or between individual network of the same type (e.g., between different brain networks). Further, we investigate if the geometry of equi-M sets may deliver a more efficient (dynamics based) description and classification tool than simple graph theoretical connectome measures. Additionally, but beyond the scope of this study, is to develop a theoretical correspondence between properties of the connectome and geometric properties of the equi-M set.

Taken together, this raises the possibility of using equi-M sets as an assessment/classification toolbox for brain functional dynamics. In this framework, each subject's connectome can be used to compute a fractal, topological signature of the subject's equi-M set (much like a finger-print). This signature (1) is based on the connectivity architecture of the brain, since the node-wise quadratic dynamics are used in conjunction with the specific connectome for each individual; (2) is designed to encode the long-term, global dynamic outcome for the network, without having to biophysically model each node. In this paper, we explore for the first time this practical possibility, by computing equi-M sets for a tractography-derived data set. We illustrate the ability of the equi-M set to differentiate between individuals and groups of individuals with different physiological profiles. We show that equi-M topological measures can outperform graph theoretical measures, by illustrating dynamic properties that connectome measures fail to capture.

\section{Methods}

\subsection{Subjects and DTI data}

Our analysis is based on tractography-derived, neural connectivity data obtained from the S1200 release of the Human Connectome Project~\cite{van2013wu}. The project released to the public domain extensive MRI, behavioral, and demographic data from a large cohort of individuals ($>1000$). For our own study, we considered a subject subgroup of $N=197$ individuals, age 22-36 years (107 male, mean age $\mu \sim 25.8$ years, and 90 female, mean age $\mu \sim 28$ years). 

\subsection{Structural network construction}
\label{structural}

Preprocessed diffusion weighted images, part of the Human Connectome Project, were used to construct structural connectomes for the subjects. Fiber tracking was done using DSI Studio with a modified FACT algorithm~\cite{yeh2013deterministic}. As a first step, data were reconstructed using generalized q-sampling imaging (GQI)~\cite{yeh2010generalized}. Diffusion weighted images were reconstructed in native space and the quantitative anisotropy (QA) for each voxel was computed. Fiber tracking was performed until 250,000 streamlines were reconstructed with angular threshold of 50o, step size of 1.25 mm, minimum length of 10mm, and maximum length of 400mm. 
Streamline counts were estimated for the parcellations schemes based on the AAL~\cite{tzourio2002automated} atlas version-1 containing 116. The AAL atlas was registered to the isotropic diffusion component (ISO) image, an output of GQI. Registering directly to the ISO image minimizes any registration issues that could arise by first registering to an individual’s T1w image. Atlas registration was conducted using FSL FLIRT~\cite{jenkinson2002improved} with default parameters. The transformation matrix obtained from the registration was applied to all regions in the AAL atlas. In case two regions were registered to the same voxel, the voxel was assigned to the region with the highest probability. In order to fully sample the fiber orientations within a voxel, tracking was repeated 250x with initiation at a random sub-voxel position generating 250 connectivity matrices per subject.

For each subject, a symmetric, non-negative, weighted structural connectivity matrix, $A$, was constructed from the connection strength based on the number of streamlines connecting two regions The final connectivity matrix included edges which were present in at least 10\% (or 25) of the 250 matrices. Connectivity matrix was normalized by dividing the number of streamlines between each two coupled regions, by the combined volumes of the two regions. For each connectome, we computed a set of graph theoretical measures, consisting of node degree centrality, eigenspectrum, clustering coefficient, betweenness centrality, eigenvalue centrality, local efficiency, numbers of 3-motifs and 4-motifs, using the Brain Connectivity Toolbox~\cite{rubinov2010complex}.



\subsection{Topological measures of asymptotic dynamics}
\label{topological}

For each individual connectome, the equi-M set was generated, using 100 iteration steps and $400 \times 400$ spatial resolution. These parameters were chosen to explore in this paper a first round of estimates. These can be easily improved in future iterations of this analysis, at higher computational cost. For each M set, we computed a set of topological measures describing its position and size. The measures are based on common topological landmarks (like the cusp and the tail), use the symmetry of the set with respect to the real axis, and the fact that all equi-M sets in our data set presented as one connected component. This represents a preliminary set of measures, meant to illustrate in principle the relationships are hand. Potential candidates for more computationally advanced measures, aimed to better capture finer topological detail are proposed for future analyses in Section~\ref{limitations}.\\

\noindent \textbf{\emph{The positions of the cusp ($\gamma$) and of the tail ($\tau$).}} The presence of the cusp is a robust, network-independent feature of the equi-M sets throughout our data set -- unlike the tail structure, which is very fragile to network modifications. However, the position $\gamma$ of the cusp along the real axis varies with the network properties, as illustrated in Figure~\ref{single_set_measures}. For a fixed equi-M set, $\gamma$ is easy to compute (as the largest coordinate reached by the boundary of the equi-M set along the real axis), and relatively robust to choosing the spatial resolution. The accuracy of the tail position $\tau$, computed as the leftmost point of the equi-M set along the real axis, is more resolution-dependent, since the tail is connected by thin filaments.\\

\noindent \textbf{\emph{The area ${\cal A}$ of the set in the complex plane}}.  Our computation algorithm provides an overestimate of the exact value of ${\cal A}$, due to the finite-iteration based approximation of the asymptotic equi-M set (some of the equi-M points retained in our set representation may in reality escape upon further iteration). Note that spatial resolution of the computation may limit detection of thin filaments; these, however, have negligible area, hence they are not expected to have a major contribution to the value of ${\cal A}$.\\

\noindent \textbf{\emph{The horizontal diameter $d_h$}}, is the distance between the cusp and the tail along the real axis. \\

\noindent \textbf{\emph{The vertical diameter $d_v$}} is the largest vertical distance between two points of the set. Our computational algorithm provides us slight with underestimates of this measure; more exact computations are problematic, due to the presence of thin filaments around the boundary.\\

\noindent With these measures at hand, we pursue two questions: (1) establishing whether there are correlations between graph theoretical measures of the connectomes and geometric measures of the equi-M sets, and (2) testing whether the topology of the equi-M set can effectively differentiate between the subjects' gender (as a proof of principle, to be potentially extended in the future to other physiological or behavioral measures). The methods used to investigate these two questions are further described below.

\subsection{Correlation analysis}

We investigate how the graph structure of the connectome (representing fixed, hardwired connectivity information, as captured in the set of graph theoretical measures in Section~\ref{structural}) reflects into the shape of the equi-M set (which encompasses how information propagates in the CQN, as represented by the set of topological measures in Section~\ref{topological}).

We computed both Pearson and Spearman correlations between the graph theoretical and the topological measures. Since the topology of the M set can be viewed as a symbolic representation of the network's long term dynamics, and since the network architecture (underlying graph) is a factor in determining these dynamics, correlations between properties of one and the other may help understand which graph theoretical properties lead to which types of long-term dynamics. Understanding this relationship at the level of simple quadratic dynamics may be of crucial importance when extrapolating to the more complex, neural dynamics that occurs in reality in these brain networks.

In addition, correlations computed \emph{within} the set of topological measures may help identify to what extent these measures are inter-related. Similarly, correlations computed \emph{within} the set of graph theoretical measures will quantify the degree of redundancy in the information provided by these measures.

\subsection{Gender based statistics}

We test if the shape of the equi-M set, can distinguish statistically between the subjects' gender. In order to make the least assumptions on our distributions, we used a Mann-Whitney nonparametric rank test (as provided by the Matlab 2020a package) to assess if each measure differs significantly ($p<0.001$) between the Male and Female subjects (the null hypothesis being that they are extracted from the same overall distribution). 



\section{Results}

\subsection{Between subject differences}
\label{between_subj}

\begin{figure}[h!]
\begin{center}
\includegraphics[width=0.9\textwidth]{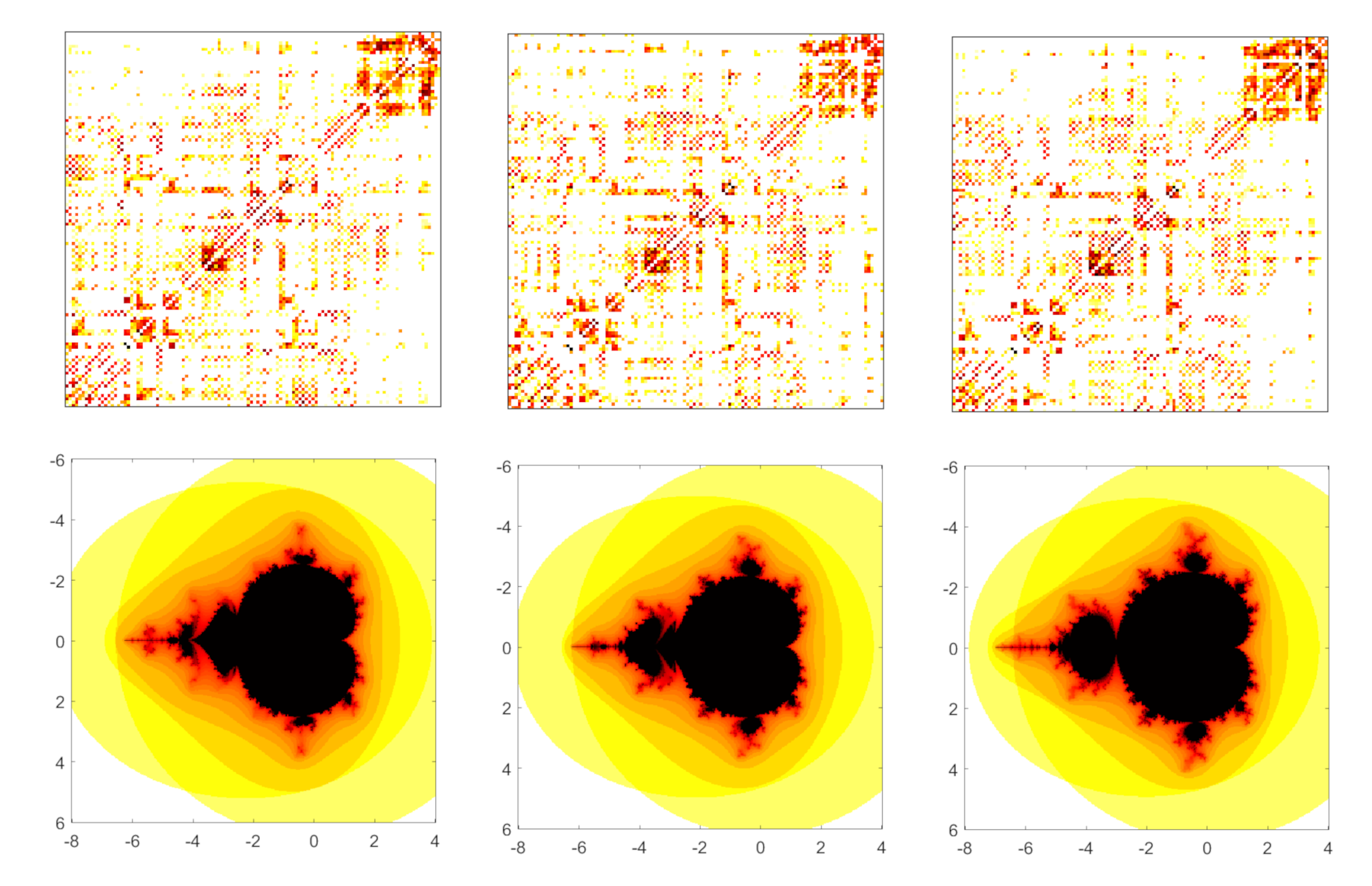}
\end{center}
\caption{\small \emph{{\bf Examples of M sets for three connectomes in our data set.} The connectomes are sparse; that is, many entries are zero, marking node pairs not being connected. In additions, many nonzero entries are very small, corresponding to weak connections. Subtle differences in the connectomes reflect into noticeable, quantifiable differences in the topology of the equi-M sets, so that each individual has a unique, distinctive equi-M set, like a personal signature, or a fingerprint encoding the CQN dynamic flow through their brain connectome.}}
\label{example_Msets}
\end{figure}

Our computation of equi-M across all 197 subjects revealed between-subject differences. Figure~\ref{example_Msets} illustrates these differences between three example individuals (with the top panels of the figure showing the subjects' data-derived connectomes, while the bottom panels show the corresponding equi-M sets).

While differences in the shapes of these three sets are undeniable, they are also subtle, especially when set against the theoretical potential for variability in shape as compared to the more dramatic differences in shape observed between the examples in Figure~\ref{M_examples}, in response to making only local changes to the network connectivity. This was true in general for the entire data set, with all the 197 equi-M sets showing  unifying geometric features (a common `signature'' to all connectomes): all equi-M set we obtained have a cusp and a tail, and as structure reminiscent of the main cardioid of the traditional Mandelbrot set. At the same time, much like fingerprints, the sets exhibit extensive variations in detail, due to subtle differences in the connectome. 

To formally test these impressions, a quantifiable assessment of equi-M set topology is needed. However, this is difficult, because the dynamics no longer capture the simple combinatorics of a single iterated map. The hyperbolic bulb structure of the traditional Mandelbrot set, which could have been helpful for classifications, no longer exists in that form for network equi-M sets. Instead, we used position and shape landmarks (see Methods section), which allow us to quantify subtle versus significant geometric differences, and to distinguish between shapes.

\begin{figure}[h!]
\begin{center}
\includegraphics[width=0.9\textwidth]{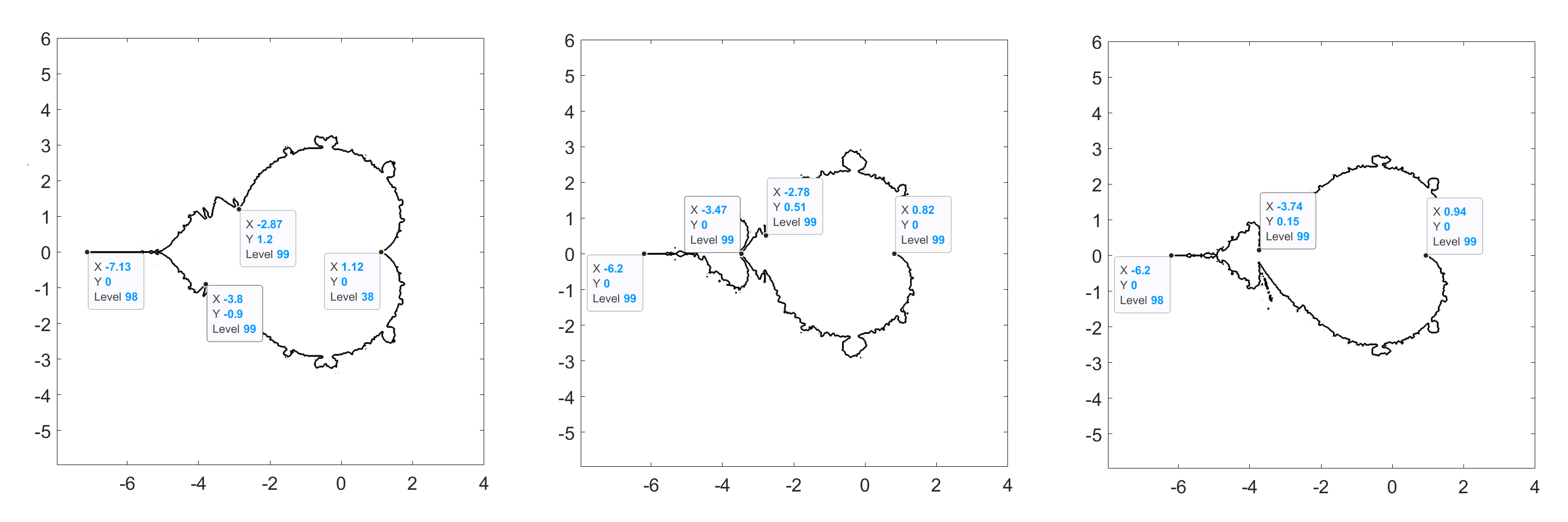}
\end{center}
\caption{\small \emph{{\bf Landmarks for Equi-M sets.} Each panel shows one DTI-generated equi-M set, with labels that facilitate computation of the measures proposed in the text: {\bf A.} $\gamma=1.12$, $\tau=-7.13$,$d_h=8.25$; {\bf B.} $\gamma=0.82$, $\tau=-6.2$, $d_h=7.02$ ; {\bf C.} $\gamma=0.94$, $\tau =-6.2$, $d_h=7.14$.}}
\label{single_set_measures}
\end{figure}

Figure~\ref{single_set_measures} illustrates these landmarks for the same three equi-M sets as those in Figure~\ref{M_examples}. Only the contours of the sets are shown here, to avoid overcrowding the panels. The set in the left panel distinguishes itself easily as the largest, with the leftmost position for tip of the tail, the rightmost cusp, and hence the largest diameter and area. However, beyond the differences in size, we can point out that the set in the middle panel has a ``pinch point'' on the real axis, at $x=-3.47$, which, if removed, breaks the set into two connected components, reminiscent of the original bulbs. In turn, the set in the right panel has a ``narrow bridge'' at $x=-3.74$, with very small diameter $2y = 0.3$ across, separating the main body from the tail. For the set on the left, the smallest vertical diameter occurs at $x=-2.87$ and is $2y = 2.4$, with no relevant separation between the body and the tail.

Pinch points and narrow bridges show future promise towards assembling a finer and more complete geometric assessment of the equi-M set signature. For the current study, we will strictly focus on using the collection of measures described in the Methods section, for all of our assessments. Based on these assessments, we investigated the graph theoretical source of the similarities and differences between sets.

\subsection{Inter-correlations between graph and topological measures}
\label{results_corr}

We want to establish if these is a well-defined correspondence between graph teoretical properties of the connectomes in the data set, and geometric properties of the corresponding equi-M sets.

\begin{figure}[h!]
\begin{center}
\includegraphics[width=0.9\textwidth]{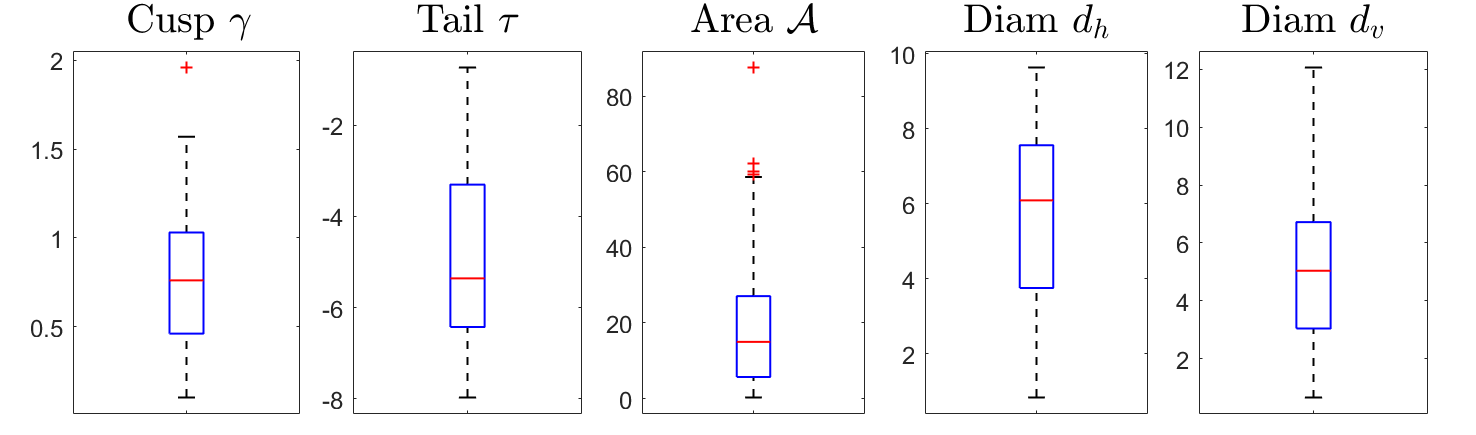}
\end{center}
\caption{\emph{\small {\bf Topological measures statistics (means $\mu$ and standard deviations $\sigma$).} Cusp position $\gamma$: $\mu = 0.77$; $\sigma = 0.36$. Tail position $\tau$: $\mu=-4.88$; $\sigma=1.87$. Area ${\cal A}$: $\mu=18.52$; $\sigma=15.29$. Horizontal diameter $d_h$: $\mu=5.65$; $\sigma=2.22$. Vertical diameter $d_v$: $\mu=5.10$; $\sigma=2.45$.}}
\label{top_stats}
\end{figure}

\begin{figure}[h!]
\begin{center}
\includegraphics[width=0.7\textwidth]{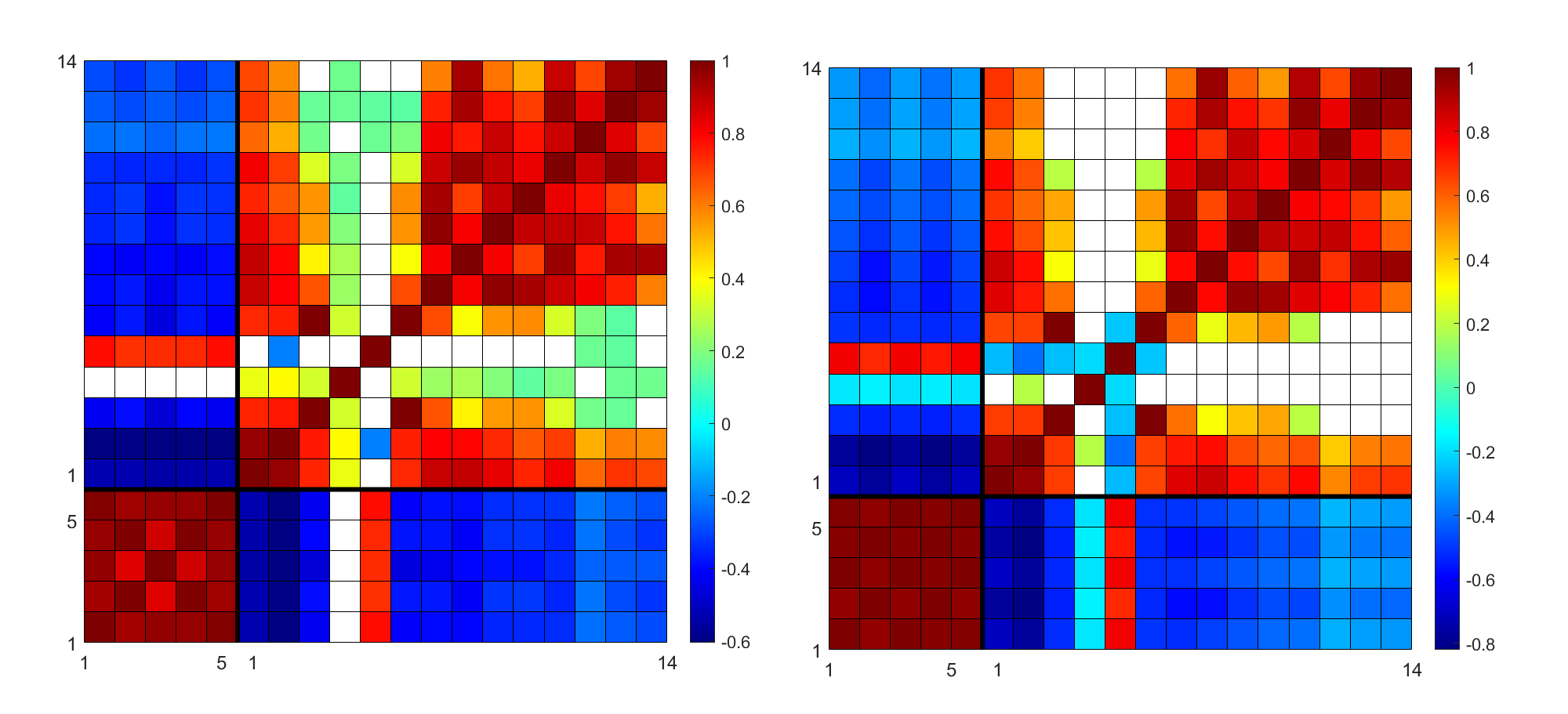}
\end{center}
\caption{\emph{\small {\bf Correlations within and between graph theoretical and topological measures.} {\bf A.}  The panel represents in color the correlation values for all pairs of measures for which the correlation value was significant (with a p-value threshold $\alpha =0.05$). The color coding for sign and strength of correlations are specified by the color bar. The pairs for which correlations were nor significant are shown in white. On both axes, topological measures are labeled 1-5 (representing the following order: 1 = cusp position $\gamma$, 2 = absolute value of the tail position $\tau$, 3 = area ${\cal A}$, 4 = diameter along real axis $d_h$, 5 = vertical diameter); then graph theoretical measures are labeled 1-13 (representing the following sequence: 1 = degree centrality; 2 = eigenspectrum; 3 = clustering coefficient; 4 = betweenness centrality; 5 = eigenvalue centrality; 6 = local efficiency; 7-8 = numbers of 3-motifs; 9-13 = numbers of 4-motifs.}}
\label{Rvals_graph_top}
\end{figure}

For each subject, we computed the set of topological measures consisting of $(\gamma, \tau, {\cal A}, d_h, d_v)$ (as described in the Methods section). The mean and standard deviation of all four measures are described in Figure~\ref{top_stats}. The correlation analysis confirmed that these measures are not independent in the context of our data. Figure~\ref{Rvals_graph_top} suggests that they are strongly correlated with each other (e.g., a position of the cusp more to the right also corresponds to a tail shifted to the left, larger diameters and a larger area).\\

For each connectome, the set of graph theoretical measures described in the Methods section were also computed.
We performed correlation analyses, investigating the inter-relationship between the topological and the graph theoretical measures. The results of both Pearson and Searman correlation analyses are shown in Figure~\ref{Rvals_graph_top}. In both analyses, we found strong correlations between most pairs (overall average p-value $p=5.9088 \times 10^{-4}$). Higher Degree Centrality, Eigenspectrum and Local Efficiency correlated significantly with smaller equi-M sets. Specifically, the cusp and tail landmarks were closer to the origin, and the area and diameters were smaller. In contrast, the Eigencentrality measure correlated positively with the size of the equi-M set (i.e., higher Eigencentrality values corresponded to larger equi-M sets). Betweenness Centrality did not show significant correlations. Interestingly, the number of motifs showed consistent significant negative correlations with the size of the equi-M set: the stronger the motifs, the smaller the topological measures. A potential explanation to this correlation between motif strength and size of the equi-M set is proposed in our previous work~\cite{radulescu2022synch}, but a complete proof is nontrivial, and still under way.

This analysis confirms that there is a tight correspondence between connectome patterns and  geometric patterns of the equi-M set. This approach can be refined with developing and using finer topological measures for the equi-M sets, as further considered in the Discussion.


\subsection{Gender based statistics}
\label{results_gender}

\begin{figure}[h!]
\begin{center}
\includegraphics[width=\textwidth]{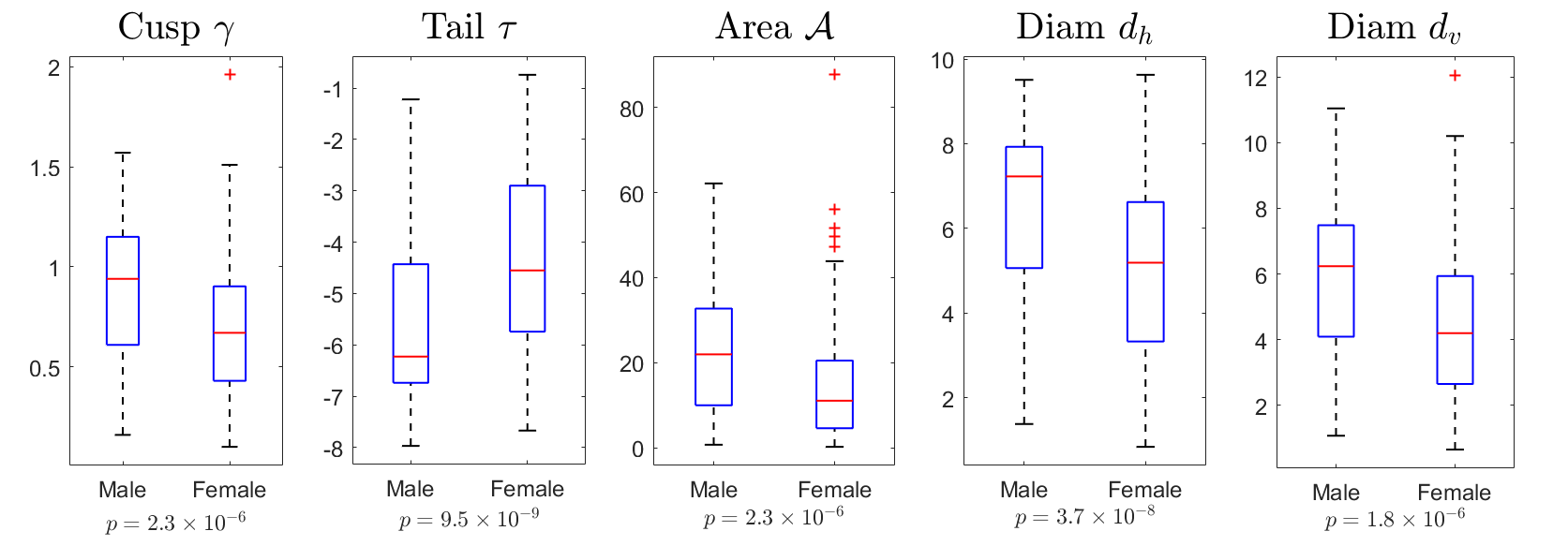}
\end{center}
\caption{\emph{\small {\bf Comparison of topological measures statistics by gender.} {\bf Male statistics.} Cusp position $\gamma$: $\mu = 0.88$; $\sigma = 0.36$. Tail position $\tau$: $\mu=-5.52$; $\sigma=1.76$. Area ${\cal A}$: $\mu=23.08$; $\sigma=16.19$. Horizontal diameter $d_h$: $\mu=6.40$; $\sigma=2.10$. Vertical diameter $d_v$: $\mu=5.85$; $\sigma=2.42$. {\bf Female statistics.} Cusp position $\gamma$: $\mu = 0.63$; $\sigma = 0.33$. Tail position $\tau$: $\mu=-4.12$; $\sigma=1.73$. Area ${\cal A}$: $\mu=13.10$; $\sigma=12.26$. Horizontal diameter $d_h$: $\mu=4.76$; $\sigma=2.04$. Vertical diameter $d_v$: $\mu=4.21$; $\sigma=2.20$. The $p$-values of the Male/Female between-group comparison (Wilcoxon ranksum test) can be found below the box corresponding to each moeasure.}}
\label{gender_boxplot}
\end{figure}

\begin{figure}[h!]
\begin{center}
\includegraphics[width=0.8\textwidth]{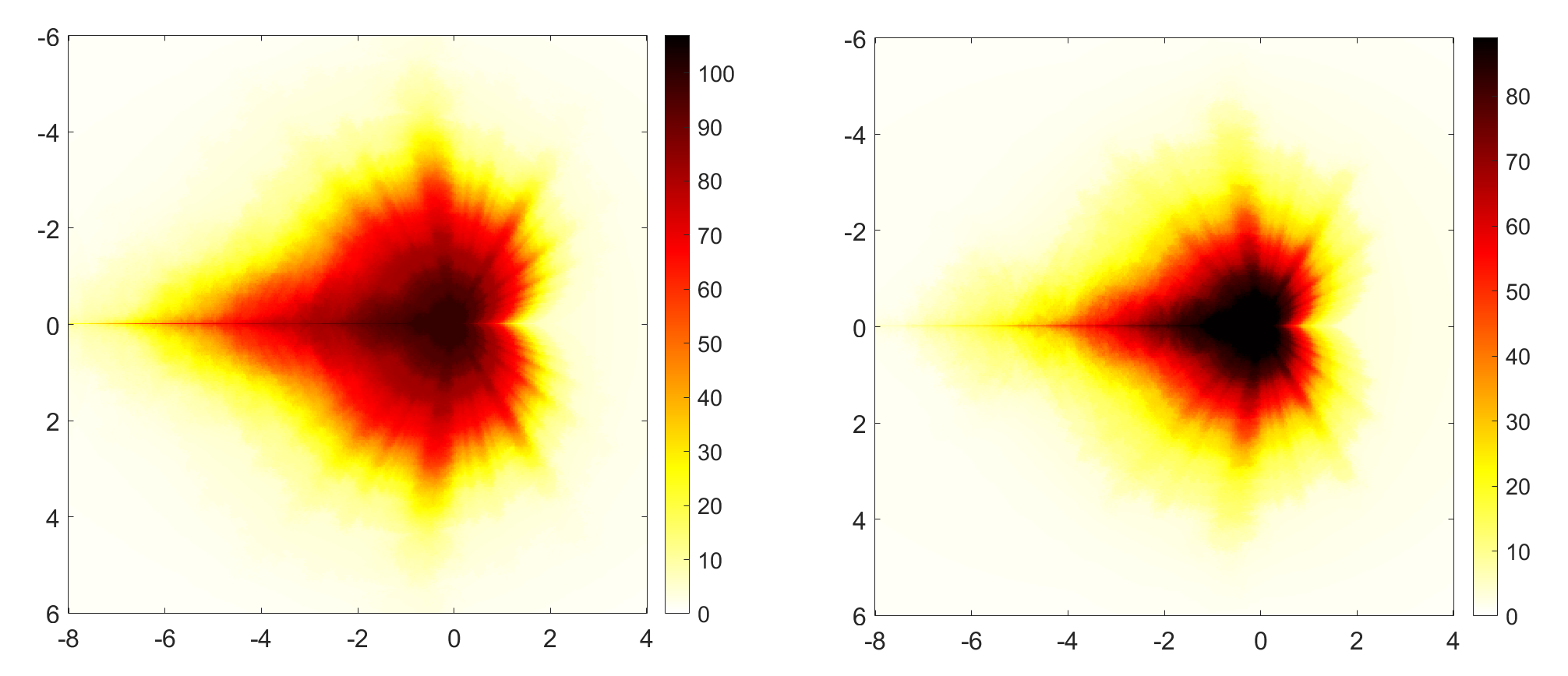}
\end{center}
\caption{\emph{\small {\bf Stochastic versions of the equi-M set} for the Male group (left panel) and for the Female group (right panel).The color for each point in these parameter squares represents for how many subjects the $c$ value corresponding to the point is in the equi-M set of the subject' network. The color map used (with lighter colors for lower numbers, and darker colors for higher numbers) is indicated for each group in the color bar. Each figure panel presents a stochastic representation of the equi-M sets in each subject group (``frequency plot''), constructed as follows: to each point in the complex plane, we associated the number of subjects in the group for which that point is in the equi-M set. Then we used a color map to represent this as a 2-dimensional plot (with darker colors corresponding to more subjects, and lighter colors to fewer subjects, as specified in the color bar). This conveys visually ideas such as for example, that the sets are smaller and present less size variability within the Female group (the colored region is smaller than the corresponding one in the Male group, and have less variability (the halo of transitional colors is thinner than that for Males).}}
\label{gender_diffs}
\end{figure}

Our sample population consisted of healthy young adult male and females. To test the utility of our analytical approach, we investigated if our framework is able to identify gender differences. We found that our model is able to capture previously unidetified and nuanced differences between the Male and Female groups in our sample.  
Specifically, the shapes of the equi-M sets correlated with the sex of the subjects. These gender differences for the topology measures are described in Figure~\ref{gender_boxplot}. The equi-M sets corresponding to the Male subgroup have significantly larger cusp coordinates, longer tails, larger areas, horizontal and vertical diameters than the female counterparts. Figure~\ref{gender_diffs} provides an illustration of the between-gender differences in topology stemming from differences in connectivity.

\begin{figure}[h!]
\begin{center}
\includegraphics[width=0.7\textwidth]{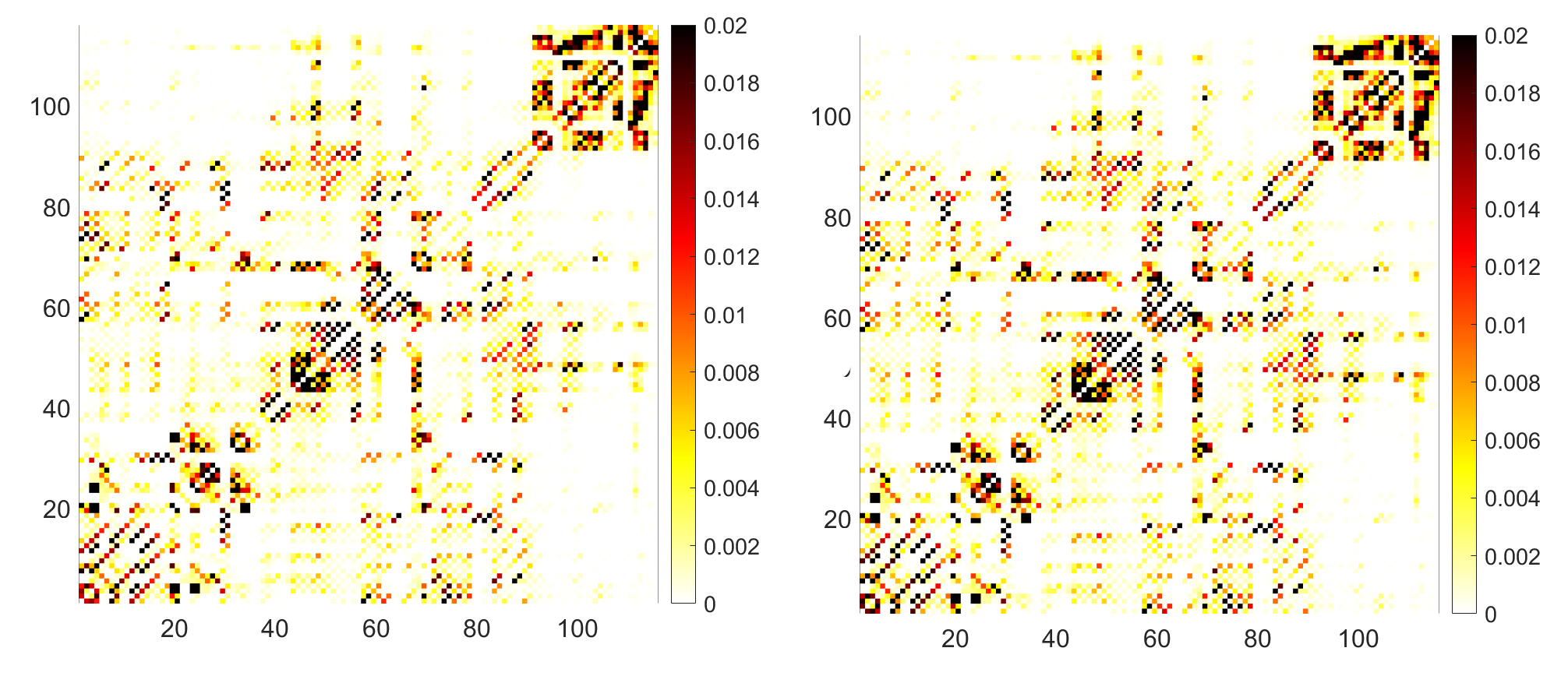}
\includegraphics[width=0.7\textwidth]{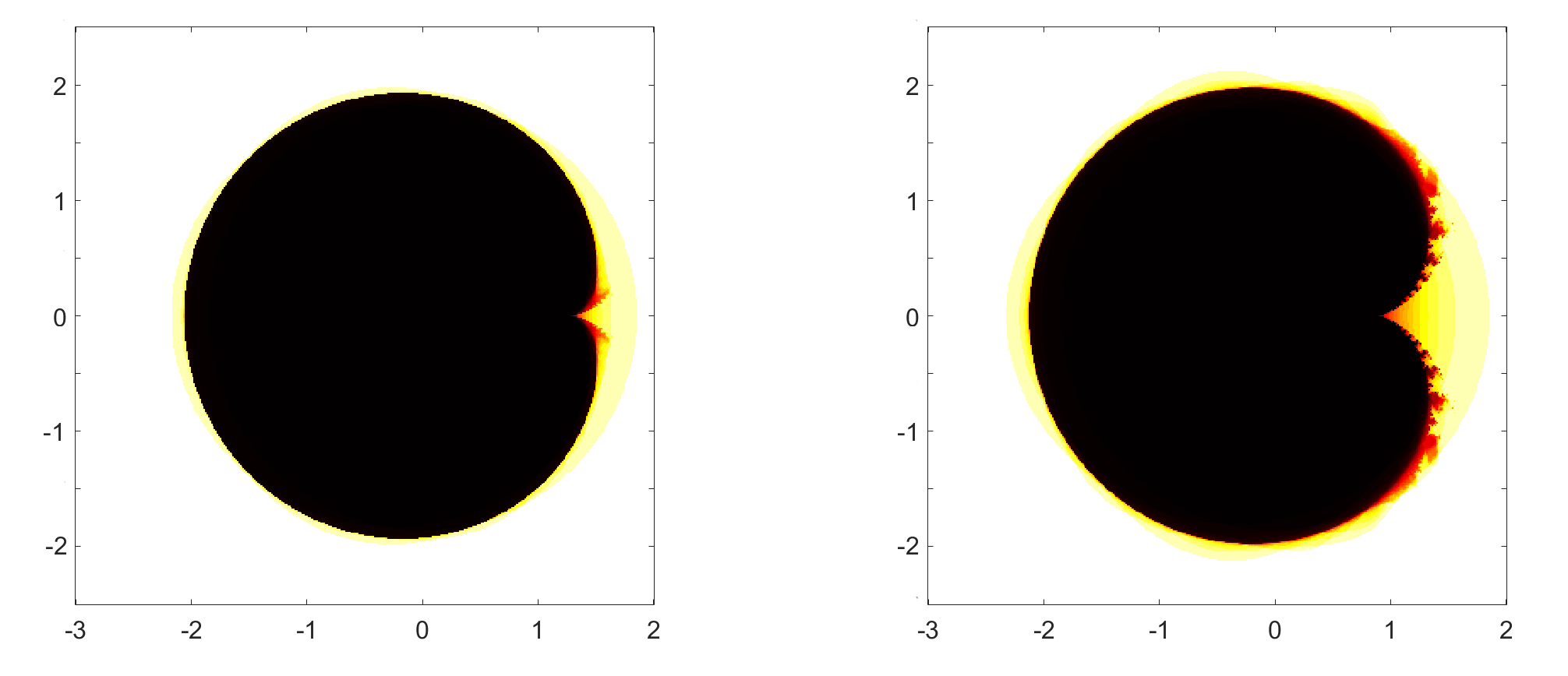}
\end{center}
\caption{\emph{\small {\bf Prototypical connectomes and equi-M sets for the Male and Female groups.}  {\bf Top.} Prototypical connectomes for Male subjects (left) and Female subjects (right), obtained by averaging the individual connectomes for the respective subject groups. {\bf Bottom.} Prototypical equi-M sets for Male (left) and Female subjects (right), obtained from the prototypical connectomes for the respective groups.}}
\label{gender_proto}
\end{figure}

Further, in Figure~\ref{gender_proto}, we highlight how topology outperforms graph theory in identifying gender differences. We computed two group “mean” (or “prototypical”) connectomes, by averaging the individual connectomes over each subject group (Males and Females, respectively). The Male and Female prototype connectomes are shown in Figure~\ref{gender_proto} (top row). We then computed the prototypical Male and respectively Female equi-M sets from these two connectomes, as illustrated in Figure~\ref{gender_proto} (bottom row). The two prototypical connectomes are hardly distinguishable with the naked eye; through averaging significant individual detail in connectivity patterns was averaged out, resulting in two almost identical connectomes. However, these seemingly unnoticeable differences were enough to produce visibly different equi-M prototypical sets for Males and Females (with a deeper cusp, and wider and more pronounced detail around the cusp in the set corresponding to the Female group than in the set corresponding to Males).

\section{Discussion}

Here we show for the first time how the theory of complex quadratic networks (CQNs) can be applied to improve our understanding of natural networks. We chose to illustrate the principles in a data set consisting of tractography-derived brain networks. Our study supports a few important points. 

First, we showed that, as one might expect, the geometry of the equi-M set is to a large extent encoded in measurable graph theoretical properties of the connectome. Information on the size and the distribution of the weights in the network is crucial to producing the topology of the connectomes observed empirically. Crucially, some of the unifying and the distinguishing features between the tractography-generated equi-M sets seem to be driven by features in the connectome that might not be captured by traditional graph theoretical measures (see for example the prototyical sets in Figure~\ref{gender_proto}). To obtain a better understanding of how the structural and dynamic aspects are tied, one has to (1) better tailor the connectome measures and (2) refine the topological assessment of equi-M set (as further proposed in the section on Future Work).

Second, we explored the possibility of using the equi-M set as a classification and prediction tool. This is an interesting and promising direction of investigation, since a classification instrument which uses brain dynamics is one step closer to capturing the subjects' physiology and behavior than classifiers based entirely on the hardwired structure (i.e., the connectome).
A very important step in this direction was to show that classifications based on equi-M set topology can in fact outperform classifications using graph theoretical measures. In this paper, we give a simple illustration of this potential, by showing that the topology of the equi-M set can efficiently capture differences in dynamics between Male vs. Female prototypical connectomes with otherwise seemingly indistinguishable architecture. In future work, we plan to further explore the ability of the M set to capture physiological and behavioral markers, and its potential as a predictor of these aspects.

Overall, while previous work has generally focused on formulating and addressing mathematically some of the rich dynamics in CQNs~\cite{radulescu2017real,radulescu2019asymptotic}, the results of this study support the seemingly intriguing idea that CQNs also have very practical applications. Our work in CQN aligns in spirit with an emerging recent effort to use network models with simplified, tractable node dynamics (e.g., Threshold Linear Networks~\cite{curto2016pattern,parmelee2022core}) to mathematically tie complex network structure with coupled dynamics. While at the moment these models are still few and far between, especially in computational neuroscience, there is increasing evidence that they can be used successfully to understand and predict natural network dynamics, with sharper clarity than more complex models. Some of the limitations of our particular model, as well as a few potential research directions emerging are described in the next section.

\subsection{Limitations and future work}
\label{limitations}

The structural connectome has the intrinsic limitation of only capturing the physical connections, but not to what extent these are effectively used, or contribute to the coupled dynamics at the moment of the scan. In fact, it is likely that some of the weak connections between nodes may be typically silent, without much cost for the brain. In a parallel study, we showed that weak connections contribute crucially to CQN dynamics and synchronization. Hence access to the \emph{functional} connectome seems to be of utmost importance when aiming to understand the network dynamics. Direction and signature of connections (excitatory versus inhibitory) were also found in our preliminary explorations to be crucial to the emerging dynamics, and down the line to the brain's function and observed behavior. Our immediate next step is to redo the current analysis on functional data from the same 197 subjects, using a matrix of effective, signed, directed connections.

One intrinsic limitation comes from the fact that the methods are still developing (hence their applicability and our understanding of their significance will evolve with the theoretical and methodological progress). For example, at this point we have only computed a network escape radius for certain network conditions, which are not necessarily satisfied by all our data-generated networks. What this implies practically is that orbits that get large over the first 50 computed iterations are not automatically guaranteed to escape, hence our numerically-generated plots mare not guaranteed to be reliable representations of the true equi-M sets. More general escape radius theoretical results, together with improved, higher-resolution numerical simulations, will permit better estimation of topological measures for equi-M sets.

\begin{figure}[h!]
\begin{center}
\includegraphics[width=0.65\textwidth]{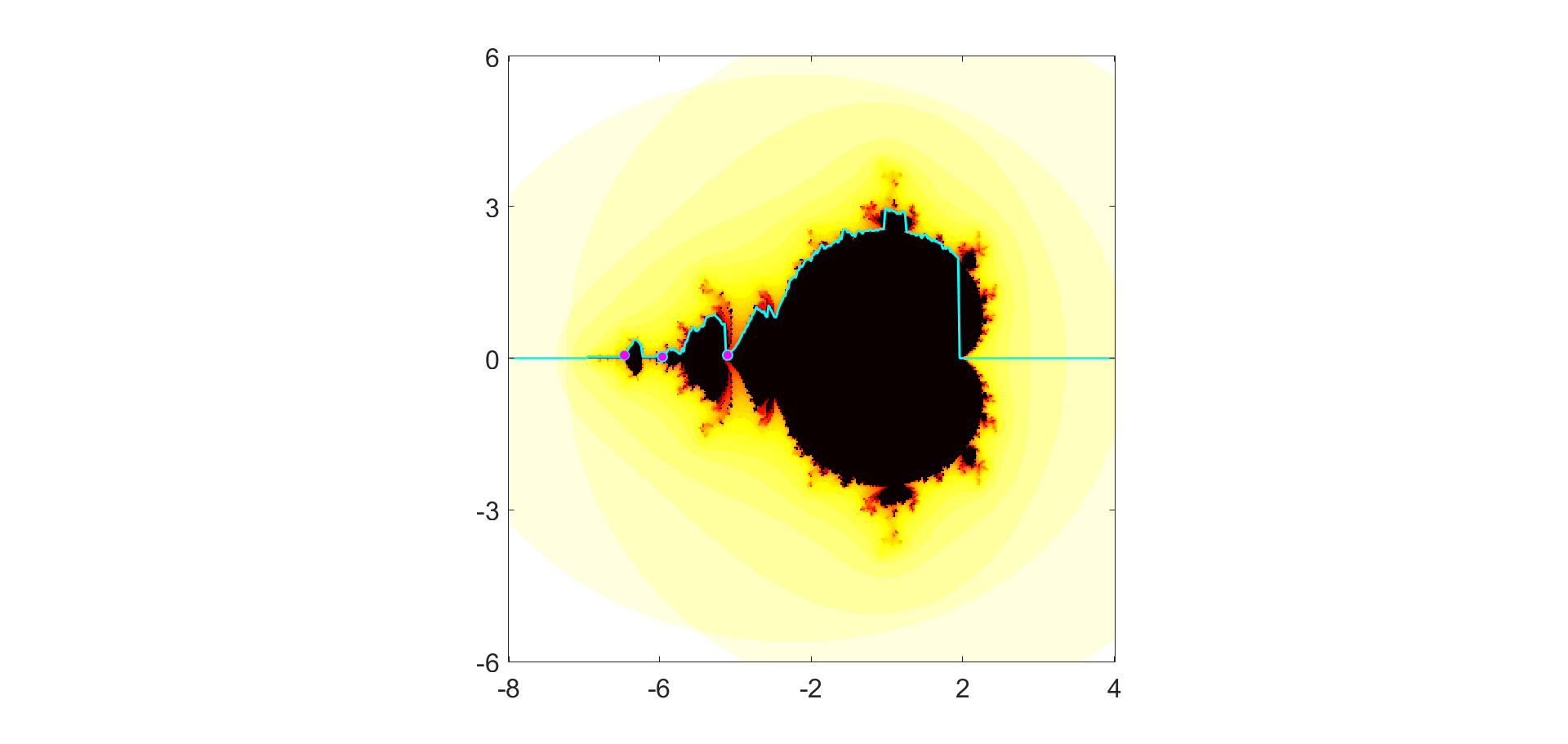}
\end{center}
\caption{\small \emph{{\bf Counting pinch points for a DTI-generated equi-M set.} The algorithm detects the pinch points (purple) points where the M set narrows out vertically to a point on the x-axis (virtually separating the connected component obtained for the equi-M set after removing the x-axis).}}
\label{top_Mset}
\end{figure}

A primary aim of our continuing work on equi-M sets is to develop additional assessment measures of the equi-M set topology. Based on our first observations, we propose two potential directions, which are expected to be able to efficiently differentiate between equi-M shapes. One goal is to capture the \emph{vertical narrowing and pinching}. The traditional Mandelbrot set has a hyperbolic bulb structure. In particular, the bulbs that are centered along the real axis are tangent to each other along the axis at what we will call ``pinch points'' (because their removal would break the set into connected components). While for equi-M sets the bulb structure is not preserved, some of the pinch points persist -- either in this form, or just as a narrowing of the set (which we call ``narrow bridge''), delimiting ``pseudo-bulbs.'' We will use the \textbf{\emph{number of pinch points $p$}} as a measure of bulb structure preservation. We expect these to be easier to compute than the maximal vertical diameter in general, since the narrowing points of the sets are not typically associated with filaments (similarly with the seahorse valley in the traditional Mandelbrot set). An alternative measure, related to the presence of narrow bridges is the entropy of the equi-M boundary, as described in Figure~\ref{top_Mset} (since it captures the regularity of the vertical variations between high and low points along the top boundary of the set). 

One limiting factor of this analysis stems from the fact that all topological measures computed describe rather simplistic properties related to the size and position of the equi-M set, and are all strongly correlated. More diverse and subtle measures (representing properties such as fractal dimension of the boundary, number of pinch points and pseudo-bulbs) may better and more thoroughly underline the distinctions between the effects of different graph structures on the dynamics. These concepts will be implemented in future work.

A longer-term future direction is to unify the results into a more statistically comprehensive approach. In this framework, one would be able to select, for a given set of connectome data, a collection of topological measures that optimally reflect the differences in the equi-M sets, as well as their relationship with the graph architecture. This can then be used as a \emph{topological profiling toolbox} (TPT) for the data-set at hand. 

The TPT can be used to produce from the connectome a ``dynamic brain signature,'' encoding the efficiency in each individual's brain regulation. The overarching goal is working on developing TPT-like lookup charts towards a neurobiologically-based diagnostic instrument that can be used by clinicians to optimize classification, prediction and intervention techniques.\\

 In future work, we plan to also compare the performance of equi-M set classifications with those delivered by realistic neural oscillator models (e.g., with Wilson-Cowan or Hudgkin-Huxley dynamics on the network nodes).
 
 We also plan to investigate whether there is indeed a common ``signature'' specific to equi-M sets of brain tractography, and if shapes corresponding to other types of natural networks have their own, different signatures.

\bibliographystyle{plain}
\bibliography{references}

\end{document}